\documentclass[prd,aps,tightenlines,twocolumn]{revtex4}
\usepackage{graphicx,color}
\usepackage{amssymb}

\newcommand{\bear}{\begin{array}}  \newcommand{\eear}{\end{array}}
\newcommand{\bea}{\begin{eqnarray}}  \newcommand{\eea}{\end{eqnarray}}
\newcommand{\beq}{\begin{equation}}  \newcommand{\eeq}{\end{equation}}
\newcommand{\bef}{\begin{figure}}  \newcommand{\eef}{\end{figure}}
\newcommand{\bec}{\begin{center}}  \newcommand{\eec}{\end{center}}
\newcommand{\non}{\nonumber}  
\newcommand{\lmk}{\left(}  \newcommand{\rmk}{\right)}

\newcommand{\ds}{\displaystyle}
\newcommand{\be}{\begin{eqnarray}}
\newcommand{\ee}{\end{eqnarray}}

\def\EQ#1{Eq.~(\ref{#1})}

\begin{document}

\title{Q-ball Instability due to $U(1)$ Breaking}
\author{Masahiro Kawasaki}
\author{Kenichiro Konya}
\author{Fuminobu Takahashi}
\affiliation{Institute for Cosmic Ray Research, 
University of Tokyo, Kashiwa 277-8582, Japan}

\date{\today}

\begin{abstract}
Q-ball is a non-topological soliton whose stability is ensured by
global $U(1)$ symmetry. We study a Q-ball which arises in the
Affleck-Dine mechanism for baryogenesis and consider its possible
instability due to $U(1)$ breaking term ($A$-term) indispensable for
successful baryogenesis.  It is found that the instability destroys
the Q-ball if its growth rate exceeds inverse of the typical
relaxation time scale of the Q-ball.  However, the instability is not
so strong as it obstructs the cosmological formation of the Q-balls.

\end{abstract}

\maketitle

\section{Introduction}

The Q-ball is a non-topological soliton that arises in scalar field
theory with some global $U(1)$ symmetry~\cite{Coleman:1985ki}. The
Q-ball solution naturally exists in the minimal supersymmetric (SUSY)
standard model
(MSSM)~\cite{Kusenko:1997zq,Dvali:1997qv,Enqvist:1997si,Enqvist:1998en},
especially in the context of the Affleck-Dine (AD)
baryogenesis~\cite{Affleck:1984fy} where the MSSM flat directions play
important roles in baryon number generation.  In this case the Q-ball
consists of squarks and sleptons, therefore carrying baryonic and/or
leptonic charges.  It was shown that Q-balls with large baryon number
are actually produced in the early universe in
Refs.~\cite{Kasuya:1999wu,Kasuya:2000wx}.  Furthermore, a Q-ball with
large baryon number is stable against decay into protons in theories
based on the gauge-mediated SUSY breaking.  Therefore, such stable
Q-ball can be a promising candidate for dark matter as well as the
source of the baryon number of the universe, which makes the Q-balls
very
attractive~\cite{Kusenko:1997si,Kasuya:2000sc,Kasuya:2001hg,Kasuya:2003yr}.
Q-balls in the gravity-mediated SUSY breaking models, on the other
hand, are unstable and decay into the standard particles, producing
the lightest SUSY particles (LSPs). Then it is possible that the
Q-balls account for the dark matter ($=$LSPs) and the baryon asymmetry
of the universe simultaneously~\cite{Enqvist:1998en}.  For further
applications of Q-balls and their variant~\cite{Kasuya:2002zs}, see
e.g.
Refs.~\cite{Kasuya:2001tp,Kawasaki:2002hq,Ichikawa:2004pb,Kawasaki:2004th}.

In the AD mechanism we need the $U(1)_{B(L)}$
breaking terms for successful generation of the baryon
(lepton) number. Here $U(1)_{B(L)}$ is the global symmetry 
associated with baryon (lepton) number, and in the following we drop
the subscript $B(L)$, since the distinction makes no difference to
the following discussion.
 If Q-balls are not formed,
the $U(1)$ breaking terms can be neglected soon after the baryon number
is generated. This is because the cosmic expansion decreases
the amplitude of the AD field and the $U(1)$ breaking terms
become much smaller than the $U(1)$ conserving ones. 
However, once the Q-balls are formed, since the amplitude of the AD 
field inside the Q-ball is fixed it becomes nontrivial 
whether the $U(1)$ breaking terms are truly negligible or not.
Therefore, in the present paper, we study the possible instability
due to the $U(1)$ breaking A-terms and their effect on the 
Q-ball stability. It is found that the A-term induces 
instability similar to that of the parametric resonance and it upsets 
the stability of the Q-ball if the growth rate of the induced 
instability is larger than the inverse of the relaxation time 
of the Q-ball configuration. However, in the realistic cosmological
situation, the A-term inside the Q-ball is not so strong to cause 
the strong instability, hence the previous studies
which assumed the Q-ball stability remain valid.

\section{Linear  Analysis on Instabilities}

First let us consider the instabilities due to $U(1)$ breaking term in
the homogeneous background. To this end we perform linear analysis
assuming small perturbations. The potential of the AD field $\Phi$ is
written as
\begin{eqnarray}
  V(\Phi)  & = & m_{\Phi}^{2} \left(1 
    + K \log\left(\frac{|\Phi|^{2}}{M^{2}}\right)\right)|
    \Phi|^{2}  \nonumber \\
    & & + A m_{3/2}\left(\frac{\Phi^{d}}{dM_{*}^{d-3}} + h.c.\right)+ \frac{|\Phi|^{2d-2}}{M_{*}^{2d-6}},
    \label{eq:potential}
\end{eqnarray}
where $m_{\Phi}$ is the mass of the AD field, $K$ a numerical
coefficient of the one-loop correction, $m_{3/2}$ the gravitino mass,
$M$ the renormalization scale, and $M_{*}$ some cut-off scale for the
nonrenormalizable operator.  Also $A$ is assumed to be a $O(1)$ real
parameter, and $K$ is assumed to be negative.  One can see that the
second term (called A-term) breaks the $U(1)$ symmetry.  In the
following argument we can safely neglect the last term since the
amplitude of the AD filed is relatively small.

In order to study the instabilities of the AD field we first divide
the AD field into homogeneous part and fluctuation: $\Phi =
\bar{\Phi}+\delta \Phi$. The equations of motion are
\bea
\label{eq:eom_homo}
&&\ddot{\bar{\Phi}} + 3 H \dot{\bar{\Phi}} + m_{\Phi}^2 
\lmk1+K+K \log\lmk\frac{|\bar{\Phi}|^2}{M^2} \rmk \rmk \bar{\Phi} \non\\
&&~~~+ A m_{3/2} \frac{{\bar{\Phi}}^*{}^{d-1}}{M_*^{d-3}} = 0,
\eea
for the homogeneous mode, and
\bea
\label{eq:eom_fluc}
&&\delta \ddot{ \Phi} + 3 H \delta \dot{\Phi} + \frac{k^2}{a^2} \delta \Phi
+ K m_{\Phi}^2\bar{\Phi} \lmk \frac{\delta \Phi}{\bar{\Phi}} + \frac{\delta \Phi^*}{\bar{\Phi}^*}\rmk \non\\
&&~~~~+m_{\Phi}^2  \lmk1+K+K \log\lmk\frac{|\bar{\Phi}|^2}{M^2} \rmk \rmk  \delta \Phi \non\\
&&~~~~+A \lmk d-1 \rmk m_{3/2} \frac{{\bar{\Phi}}^*{}^{d-2}}{M_*^{d-3}} \delta \Phi^*= 0,
\eea
for the fluctuation with the wavenumber $k$ in the momentum space.
Although we have introduced the Hubble parameter $H$ and the scale
factor $a$ in the above equations, we will neglect the cosmic
expansion for the moment.

\begin{figure}[t]
\begin{center}
\includegraphics[width=8.5cm]{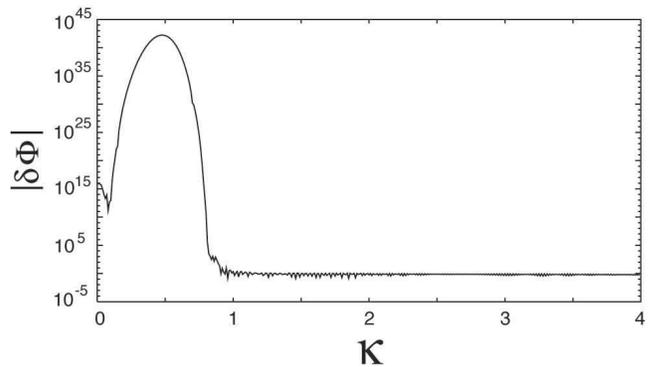}
\caption{Instability bands due to the A-term for $d=4$,
$\xi = 0.1$ and circular orbit. }
\label{fig:d4inst}
\end{center}
\end{figure}

\begin{figure}[t]
\begin{center}
\includegraphics[width=8.5cm]{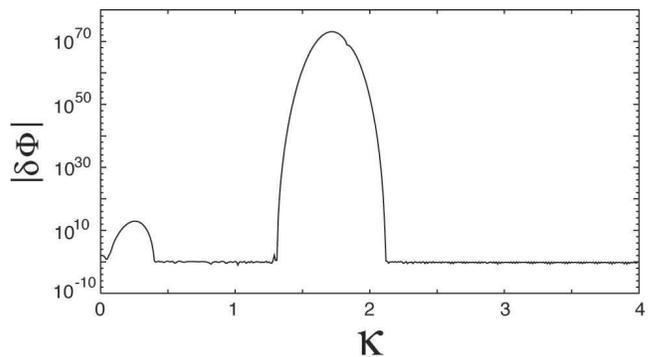}
\caption{Same as Fig.~\ref{fig:d4inst}  but $d=6$.}
\label{fig:d6inst}
\end{center}
\end{figure}

In order to parametrize the strength of the instability, let us define
$\xi$ as the ratio of the $A$-term to the mass term:
\begin{equation}
    \xi \equiv 
 2\frac{|A|m_{3/2}\Phi_{0}^{d-2}}{d m_{\Phi}^{2} M_{*}^{d-3}},
 \label{eq:def_of_xi}
\end{equation}
where $\Phi_0$ is the maximal magnitude of $\Phi$ during the course of
the oscillation in the limit of $\xi \rightarrow 0$ and we will set
$M=\Phi_0$.  Note that the amplitude of the oscillation in the limit
of $\xi \rightarrow 0$ is constant if we neglect the cosmic
expansion. We have numerically solved Eqs.~(\ref{eq:eom_homo}) and
(\ref{eq:eom_fluc}) by decomposing $\bar\Phi$ and $\delta \Phi$ into
their real and imaginary components:
\bea
\bar\Phi &=& \frac{\phi_1+ i \phi_2}{\sqrt{2}}, \non\\
\delta \Phi &=& \frac{\delta \phi_1+ i \delta \phi_2}{\sqrt{2}}.
\eea
For convenience we also define the polar decomposition: $\bar \Phi =
\phi \,e^{i \theta}/\sqrt{2}$.  We have normalized the dimensionless
variables as $\varphi_i = \phi_i/m_{\Phi}$, $\delta \varphi_i = \delta
\phi_i/m_{\Phi} ~~(i=1,2)$, $\kappa= k/m_{\Phi}$, $\tau=m_{\Phi} t$,
and $\chi_{i} =m_{\Phi} x_{i} ~ (i=1,\cdots,D)$, where $D$ is the
spatial dimension.  Furthermore we set $K=0$ in order to concentrate
on the instability that originates from the A-term.
Fig.\ref{fig:d4inst} shows the instability bands for $d=4$ and $\xi =
0.1$ with the initial condition $\theta = \pi/8$ and $\theta' = 1$,
where the prime denotes the differentiation with respect to $\tau$.
Notice that with the above initial values the orbit of the homogeneous
mode is close to a circle in the $(\phi_1,\phi_2)$ plane.  From the
figure one can see that there exist instability bands similar to those
of the parametric resonance; the strongest instability band is located
at $\kappa = 0 \sim 0.8$ and several weaker ones at shorter
wavelengths.  We also show the result for $d=6$ and $\xi=0.1$ with the
initial condition $\theta = \pi/12$ and $\theta' = 1$ in
Fig.\ref{fig:d6inst}.  In this case the strongest instability is at
larger $\kappa$ around $1.7$.

\begin{figure}[t]
\begin{center}
\includegraphics[width=8.5cm]{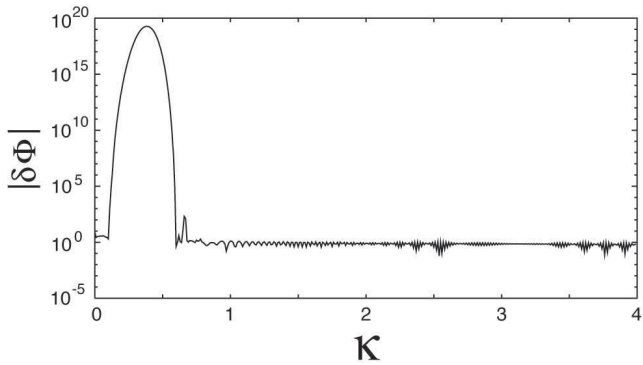}
\caption{Same as Fig.~\ref{fig:d4inst} but the elliptic orbit.}
\label{fig:d4ell}
\end{center}
\end{figure}

\begin{figure}[t]
\begin{center}
\includegraphics[width=8.5cm]{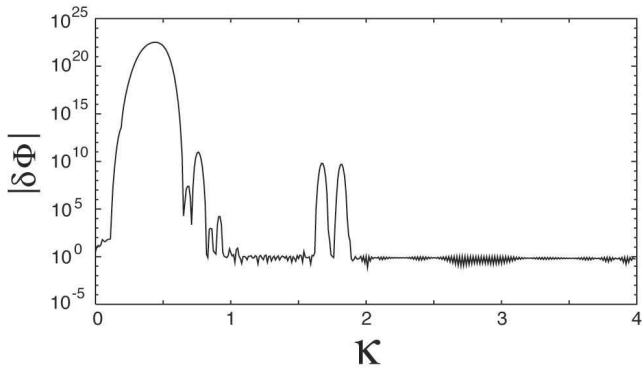}
\caption{Same as Fig.~\ref{fig:d6inst} but the elliptic orbit.}
\label{fig:d6ell}
\end{center}
\end{figure}

In Figs.~\ref{fig:d4ell} and \ref{fig:d6ell} the instability bands are
shown for the elliptic orbit of the AD field.  We have chosen the
initial conditions as $\theta = \pi/8$ and $\theta' = 0.01$ for $d=4$
and $\theta = \pi/12$ and $\theta' = 0.01$ for $d=6$, respectively.
One can see that the instability bands are located at different
wavenumber, and that the instabilities are weaker and bands narrower,
compared to the cases of the circular orbit.

\begin{figure}[t]
\begin{center}
\includegraphics[width=8.5cm]{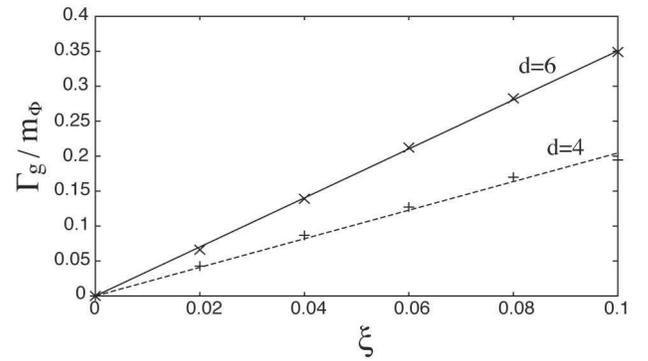}
\caption{$\xi$-dependence of the growth rate  
of the most amplified mode, $\Gamma_{g}$. 
The $+$'s and $\times$'s are the numerical results for $d=4$ and $d=6$,
respectively. 
The solid line denotes 
$\Gamma_{g}/m_{\Phi} \simeq 3.5\xi$, while the dotted one denotes
$\Gamma_{g}/m_{\Phi} \simeq 2.0\xi$. 
}
\label{fig:grow}
\end{center}
\end{figure}

In order to express the strength of the strongest instability in the case of the circular orbit, we
introduce the growth rate $\Gamma_g$;  the fluctuation
at the peak of the strongest instability band grows as $ \propto \exp(\Gamma_g t)$. 
The growth rate depends on $\xi$ and we find that
$\Gamma_{g}$ is proportional to $\xi$ as shown in Fig.\ref{fig:grow}.
Thus we obtain
\begin{eqnarray}
  \Gamma_{g} / m_{\Phi} = \gamma_{d}\, \xi\,  ,
  \label{eq:grow}
\end{eqnarray}
with $\gamma_{4} \approx 2.0$ and $\gamma_{6} \approx 3.5$. 

Before closing this section, let us consider the effect of the cosmic
expansion on the instabilities. When the oscillation of the AD field
starts, the Hubble parameter $H$ is comparable to $m_{\Phi}$. Hence
the growth rate of the A-term instability is comparable to $H$ only if
$\xi\sim O(1)$. However, since the amplitude of the AD field decreases
as $a^{-3/2}$, $\xi$ quickly becomes smaller ($\propto a^{-3(d-2)/2}$)
and the growth rate is soon overcome by the cosmic
expansion. Therefore, it seems that the A-term instability does not
play an important role at least in the Q-ball formation. However, it
is still nontrivial how the instability affects the evolution of
Q-balls after they are formed, since the amplitude inside them is
fixed.

\section{Instabilities inside Q-ball}

In the previous section we have found that the A-term causes the
instabilities in the homogeneous motion of the AD field. Next let us
investigate the instabilities due to the A-term inside the
Q-ball. Without the A-term, the Q-ball solution for the potential
(\ref{eq:potential}) is obtained with use of the Gaussian
ansatz~\cite{Enqvist:1998en}:
\begin{eqnarray}
   \label{eq:Gaussian}
  \Phi(t,r) & = & 
  \frac{1}{\sqrt{2}}e^{i\omega t}\phi(r),  \\
  \phi(r) & = & \phi(0)e^{-\frac{r^{2}}{R^{2}}},\\
\end{eqnarray}
where the Q-ball radius $R$ and angular velocity $\omega$ are given by
\begin{eqnarray}
\label{eq:qball_param}
  R^{2} & = & \frac{2}{m_{\Phi}^{2}|K|}, \\
  \omega^{2} & = & m_{\Phi}^{2}(1+2|K|).
\end{eqnarray}
To study the instabilities due to the A-term inside the Q-ball, we have numerically
solved the evolution of the AD field on one and two dimensional lattices. The
equation of motion for the potential (\ref{eq:potential}) is 
\begin{eqnarray}
&&  \ddot{\Phi} - \nabla^{2}\Phi + m^{2}_{\Phi}
  \left(1+K+K\log\left(\frac{|\Phi|^{2}}{M^{2}}\right)\right) 
  \Phi 
  \nonumber \\
   & &~~+ Am_{3/2} \frac{\Phi^{* {d-1}}}{M_{*}^{d-3}} =0.
\end{eqnarray}
We have neglected the expansion of the universe since the Q-balls decouple
from the cosmic expansion once they are formed.

In the numerical calculations we have set $M =\phi(0)/\sqrt{2}$ and
investigated several values of $K$ between $-0.01$ and $-0.3$ for both
$d=4$ and $d=6$, varying $\xi$.  For the Q-ball configuration, $\xi$
is evaluated at the center of the Q-ball; $\Phi_0$ in
\EQ{eq:def_of_xi} should be interpreted as $\Phi_c$, the field value
at the center.  Also we have added small initial seed for fluctuations
of $O(10^{-5})$.  The size of the lattices is $2048$ and $200 \times
200$ for 1D and 2D, respectively.  A number of simulations with
varying values of the lattice spacing and box size have been done to
ensure that the exact values of the lattice parameters do not affect
the results.

\begin{figure}
\begin{center}
\includegraphics[width=6.0cm]{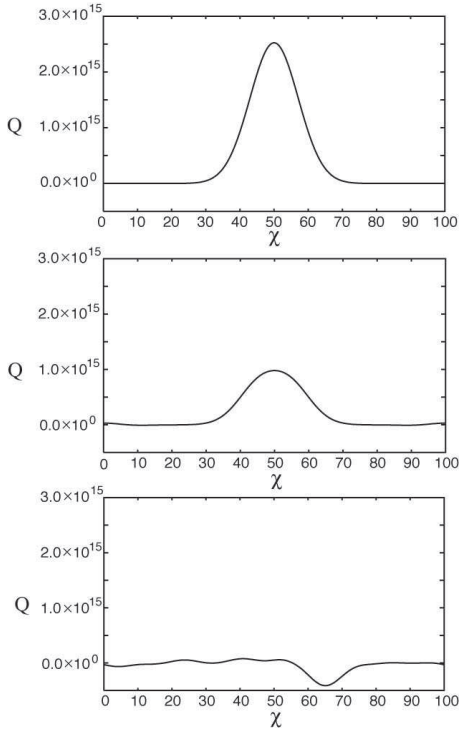}
\caption{The typical breakdown of the Q-ball due to the A-term for $d=4$ on $1+1$ lattices.
$\tau = 0$, $10^{3}$ and $10^{4}$ from top to bottom.
The adopted parameters are $K= -0.01$ and $\xi = 8\times10^{-3}$.}
\label{fig:d4break-1dim}
\end{center}
\end{figure}

\begin{figure}
\begin{center}
\includegraphics[width=6.0cm]{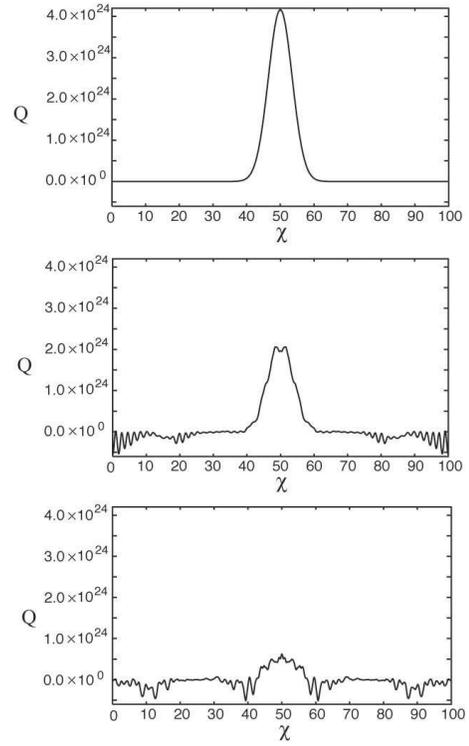}
\caption{Same as Fig.~\ref{fig:d4break-1dim} but $d=6$.
$\tau = 0$, $3.0\times10^{3}$ and $5.0\times10^{3}$ from top to bottom.
The adopted parameters are $K= -0.04$ and $\xi = 4.5\times10^{-2}$.}
\label{fig:d6break-1dim}
\end{center}
\end{figure}

\begin{figure}
\begin{center}
\includegraphics[width=5.5cm]{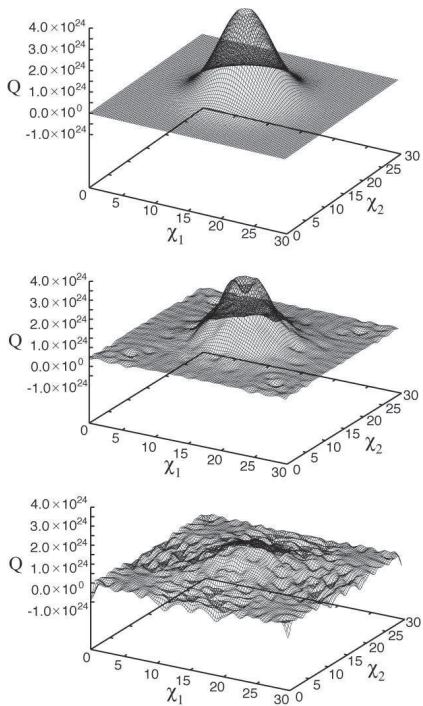}
\caption{The typical breakdown of the Q-ball due to the A-term 
for $d=6$.
$\tau = 0$, $1.0\times10^{3}$ and $1.5\times10^{3}$ from top to bottom.
The parameters we adopted are $K=-0.04$ and $\xi = 4.0\times10^{-2}$.}
\label{fig:d4break-2dim}
\end{center}
\end{figure}

Then we have found that the Q-ball breaks up for $\xi$ larger than
$O(10^{-2})$ (see Tables~\ref{tab:critical4} and \ref{tab:critical6}
for the precise values). Figs.~\ref{fig:d4break-1dim},
\ref{fig:d6break-1dim} and \ref{fig:d4break-2dim} show how the Q-balls 
are dispersed by the A-term instabilities.  It is worth noting that
the Q-ball is actually stable for small enough values of $\xi$.  The
critical values ($\xi_{c}$) above which the Q-ball becomes unstable
due to the A-term instabilities for several values of $K$ are shown in
Tables~\ref{tab:critical4} and \ref{tab:critical6}. We have followed
the evolution until $\tau = 10^4$ and then decided whether the Q-ball
configuration is lost or not on the basis of the following criterion;
if the total charge of the Q-ball is less than $10\%$ ($50\%$) of the
initial value, we judge that the Q-ball is dispersed.  We call this
criterion A (B).
  
\begin{table}[t]
\caption{The critical values $\xi_{c}$ for  $d=4$ }
\begin{center}
\begin{tabular}{|c|c|c|c|c|}
\hline
Criterion & $K=-0.01$ & $K=-0.03$ & $K=-0.1$ & $K=-0.3$ \\
\hline \hline
A & $4\times10^{-3}$ & $ 9\times10^{-3}$ & $2.8\times10^{-2}$ & $6.4\times10^{-2}$ \\
\hline
B & $3\times10^{-3}$ & $9\times10^{-3}$ & $2.6\times10^{-2}$ & $5.5\times10^{-2}$ \\
\hline
\end{tabular}
\label{tab:critical4}
\end{center}
\caption{The critical values $\xi_{c}$ for $d=6$}
\begin{center}
\begin{tabular}{|c|c|c|c|c|}
\hline
Criterion & $K=-0.01$ & $K=-0.03$ & $K=-0.1$ & $K=-0.3$ \\
\hline \hline
A & $2.2\times10^{-2}$ & $ 3.5\times10^{-2}$ & $2.3\times10^{-2}$ & $1.2\times10^{-2}$ \\
\hline
B & $2.1\times10^{-2}$ & $3.5\times10^{-2}$ & $2.3\times10^{-2}$ & $1.2\times10^{-2}$ \\
\hline
\end{tabular}
\end{center}
\label{tab:critical6}
\end{table}

The critical value $\xi_{c}$ is found to depend on the values of both
$d$ and $K$ to a certain degree. The dependence of $\xi_{c}$ on $d$
comes from the fact the instability bands have different structures
between the cases of $d=4$ and $d=6$ (see Figs.~\ref{fig:d4inst} and
\ref{fig:d6inst}).  The dependence on the other variable, $K$,
possibly comes from the following two effects; (i) the Q-ball is more
stable for larger $|K|$; (ii) the Q-ball is more tolerant to
perturbations if its charge is larger, and the charge becomes larger
for smaller $|K|$ with the amplitude $\phi(0)$ fixed because of the
dependence of the radius on $|K|$ (see \EQ{eq:qball_param}). Such a
naive reasoning might explain the behavior of $\xi_c$ especially in
Table~\ref{tab:critical6}.

Apart from the detailed dependence, can we tell a rough value of
$\xi_c$? Put another way, is $\xi_c \sim O(10^{-2})$ reasonable?  It
is conceivable that $\xi_{c}$ is determined by the competition between
the growth of the instabilities and the relaxation of the Q-ball
configuration. Since the Q-ball configuration minimizes the energy of
the system in the limit of $\xi \rightarrow 0$, the Q-ball tends to
keep its configuration for a certain range of $\xi$. It is expected
that the relaxation time scale of the Q-ball is set by the mass of the
AD field, $m_{\Phi}$, possibly multiplied by some powers of $|K|$.  On
the other hand, the growth rate calculated from Eq.~(\ref{eq:grow}) is
about $0.1m_{\Phi}$ for the critical values of $\xi$.
So, the relaxation time scale of the Q-ball should be around $10\,
m_{\phi}^{-1}$. This relaxation time scale is in an agreement with
Ref.~\cite{Multamaki:2001az} where the excited Q-balls were studied by
numerical simulations.  Therefore it is probable that the instability
due to the A-term destroy the Q-ball if its growth rate exceeds the
inverse of the typical relaxation time scale of the Q-ball.

Lastly let us comment on the realistic value of $\xi$ at the Q-ball
formation.  When the AD field starts oscillating, $\xi$ should be
order unity if $m_{3/2} \simeq m_{\Phi}$ and $A \simeq O(1)$. Then the
typical value of $\xi$ is given by $\xi \sim (\Phi_c/\Phi_{\rm
osc})^{d-2}$, where $\Phi_{osc}$ is the amplitude of the AD field at
the onset of oscillation. According to the numerical calculations (see
e.g. Ref.~\cite{Kasuya:2001hg}), $\Phi_c/\Phi_{\rm osc} \simeq 6
\times 10^{-3} (|K|/0.1)^{3/4}$. Thus the realistic value of $\xi$ is
much smaller than $\xi_c$:
\beq
\xi \sim \left\{
\bear{cc}
\ds{3 \times 10^{-5} \left(\frac{|K|}{0.1}\right)^{\frac{3}{2}} }& {\rm for~~} d=4 \\
\ds{ 10^{-9} \left(\frac{|K|}{0.1}\right)^{3}} &{\rm for~~} d=6 
\eear
\right..
\eeq
Note that  $\xi$ becomes smaller while $d$ increases.

\section{Conclusion}
In this paper we have investigated how the A-term affects the
evolution of the AD field, especially paying attention on the
stability of the Q-balls. In the linear analysis we first have found
that there exist instability bands similar to those of parametric
resonance.  The growth rate of the instabilities, however, is not so
large after the baryon asymmetry is generated.  Thus the Q-ball
formation in the expanding universe would not be disturbed by the
presence of the instabilities. Next we have studied the stability of
Q-balls and estimated the critical value of $\xi_c$, above which the
Q-balls cannot stay stable, for several values of $K$.  From our
result it is conceivable that $\xi_c$ is determined by the competition
between the growth rate and the relaxation rate of the Q-balls.  It
should be also noted that the obtained $\xi_c$ is rather large in the
realistic cosmological situations. Therefore the extensive studies on
Q-balls thus far should remain valid, since the realistic value of
$\xi$ should be smaller than $\xi_c$.  On the other hand, the
instability found in the present paper may be important when Q-balls
grow up by absorbing $U(1)$ charge such as
solitosynthesis~\cite{Griest:1989bq}.

{\it Acknowledgments.}---
  F.T.  would like to thank the Japan Society for
Promotion of Science for financial support.

\end{document}